%% file: main.tex
\documentclass[%
reprint,
superscriptaddress,
nofootinbib,
amsmath,amssymb,
aps,
]{revtex4-2}

\usepackage{booktabs}
\usepackage{graphicx}  
\usepackage{dcolumn}   
\usepackage{bm}        
\usepackage{hyperref}  
\hypersetup{
    colorlinks=true,
    linkcolor=blue,
    filecolor=magenta,
    urlcolor=cyan,
}

\def\equationautorefname~#1\null{(#1)\null} 
\def\sectionautorefname~#1\null{#1\null}    
\def\tableautorefname~#1\null{#1\null}      
\def\figureautorefname~#1\null{#1\null}     

\makeatletter
\def\p@subsection{}
\makeatother

\begin{document}


\title{Forecasting $F(Q)$ cosmology with $\Lambda$CDM background using standard sirens}


\author{Jos\'e Ferreira}
\email{jpmferreira@fc.ul.pt}
\affiliation{Instituto de Astrof\'isica e Ci\^encias do Espa\c{c}o, Faculdade de Ci\^encias da Universidade de Lisboa, Campo Grande, Edificio C8, P-1749-016, Lisboa, Portugal}
\affiliation{Departamento de F\'isica, Faculdade de Ci\^encias da Universidade de Lisboa, Campo Grande, Edificio C8, P-1749-016, Lisboa, Portugal}

\author{Tiago Barreiro}
\email{tmbarreiro@ulusofona.pt}
\affiliation{Instituto de Astrof\'isica e Ci\^encias do Espa\c{c}o, Faculdade de Ci\^encias da Universidade de Lisboa, Campo Grande, Edificio C8, P-1749-016, Lisboa, Portugal}
\affiliation{ECEO, Universidade Lus\'ofona de Humanidades e Tecnologias, Campo Grande, 376, 1749-024 Lisboa, Portugal}

\author{Jos\'e Mimoso}
\email{jpmimoso@fc.ul.pt}
\affiliation{Instituto de Astrof\'isica e Ci\^encias do Espa\c{c}o, Faculdade de Ci\^encias da Universidade de Lisboa, Campo Grande, Edificio C8, P-1749-016, Lisboa, Portugal}
\affiliation{Departamento de F\'isica, Faculdade de Ci\^encias da Universidade de Lisboa, Campo Grande, Edificio C8, P-1749-016, Lisboa, Portugal}

\author{Nelson J. Nunes}
\email{njnunes@fc.ul.pt}
\affiliation{Instituto de Astrof\'isica e Ci\^encias do Espa\c{c}o, Faculdade de Ci\^encias da Universidade de Lisboa, Campo Grande, Edificio C8, P-1749-016, Lisboa, Portugal}
\affiliation{Departamento de F\'isica, Faculdade de Ci\^encias da Universidade de Lisboa, Campo Grande, Edificio C8, P-1749-016, Lisboa, Portugal}

\date{\today}

\begin{abstract}
Forecast constraints for a Symmetric Teleparallel Gravity model with a $\Lambda$CDM background are made using forthcoming ground and space based gravitational waves observatories. A Bayesian analysis resorting to  generated mock catalogs shows that LIGO-Virgo is not expected to be able to distinguish this model from $\Lambda$CDM, while both LISA and the ET will, with the ET outperforming LISA. We also show that low redshift events are favored in order to improve the quality of the constrains.
\end{abstract}


\maketitle


\input{sections/introduction}
\input{sections/model}
\input{sections/data}
\input{sections/methodology}
\input{sections/results-discussion}
\input{sections/conclusions}
\input{sections/acknowledgments}

\bibliography{bibliography}

\end{document}

%% file: sections/introduction.tex
\section{Introduction}
\label{sec:introduction}

In \cite{GW170817} we were introduced to GW170817, the first observation of a standard siren event, a compact binary system whose merger, emitted both gravitational waves (GWs) and electromagnetic (EM) radiation. In \cite{GW190521-EM} it was also reported a possible EM counterpart to the event GW190521, introduced in \cite{GW190521}. These observations allow us to both see and "hear" extreme gravitational phenomena, providing us an unprecedented way to test our cosmological models, as standard sirens do not require a cosmic distance ladder to obtain their distances.

Although general relativity (GR) has been the most successful theory of gravity so far, it is possible that deviations from GR could explain some of the open problems in cosmology \cite{Avelino2016, MGCosmology2021}. In this standard interpretation of gravity, one sees gravity as a consequence of the curvature of spacetime. However, this is not the only way, as one can describe gravity as a consequence of a spacetime endowed also by torsion or by non-metricity. One can even manipulate these three geometrical objects such that we can build completely equivalent theories, and hence, three different interpretations, popularized as the geometrical trinity of gravity \cite{Jimenez2019a}.

The advanced Laser Interferometer Gravitational-Wave Observatory (LIGO) \cite{advLIGO} is a set of two ground based gravitational wave observatories. Together with the advanced Virgo \cite{advVirgo}, they provide the most comprehensive catalog of gravitational wave events to date. 

By placing three satellites in solar orbit, the Laser Interferometer Space Antenna (LISA) aims to be the first GW observatory in space. According to \cite{Baker2021}, it is expected to detect GWs coming from massive black hole binaries (MBHBs) with redshifts up to $z \approx 10$, which should provide us with significant cosmological insight. According to the latest proposal, presented in \cite{LISA-proposal}, the observatory is aimed to be launched as early as 2030, with a proposed lifetime of 4 years, and a possible extension to 10 years.

The Einstein Telescope (ET) is a third generation underground gravitation wave observatory, with a successful proposal, which, according to \cite{Maggiore2020}, is aimed to have its first light in 2035. Its focus is to probe for new physics in high energy events, namely, the merger of binary neutron stars (BNSs). With a significant increase in resolution, it is expected to be able to provide a glimpse of the internal structure of these massively dense stars. Although these events are expected to be detected at lower redshifts, they can still provide valuable cosmological insight.

In this paper, we will follow and describe the procedure used to generate forecast catalogs of standard sirens events for LIGO-Virgo \cite{lagos2019}, LISA \cite{Caprini2016, Tamanini2017} and the ET \cite{Belgacem2018}. We will then use these mock catalogs to constrain a specific model of Symmetric Teleparallel Gravity (STG) \cite{Jimenez2019}, that features a $\Lambda$CDM background with only one additional free parameter, where the differences arise at the perturbative level. We make use of Markov chain Monte Carlo (MCMC) methods to sample the parameter space and see whether these three observatories are expected to distinguish this model from $\Lambda$CDM in the upcoming future. An online repository complementary to this analysis is publicly available at \cite{repo}.

This work is organized as follows: first we will introduce the STG model we have considered throughout this analysis in section \autoref{sec:model}. In section \autoref{sec:datasets} we will present all of the datasets used, as well as the procedure employed to generate the standard sirens. Later, in section \autoref{sec:methodology}, we will outline the catalog selection process as well as the sampling method, section \autoref{sec:results-discussion} will show and discuss the results obtained, and finally, in section \autoref{sec:conclusions}, we will provide a general overview of the work.

%% file: sections/model.tex
\section{Cosmological Model}
\label{sec:model}

We will consider a STG model, where gravity is described by a non-metric, flat and torsion free spacetime, characterized by the action \cite{Jimenez2017}

\begin{equation}
    S = \int \sqrt{-g} \left[ -\frac{1}{16 \pi G} F(Q) + \mathcal{L}_m \right] d^4x \,,
\end{equation}
where $F(Q)$ is a generic function of the non-metricity scalar, $Q$, to be defined in equation \autoref{eq:non-metricity-scalar}, and $\mathcal{L}_m$ is the Lagrangian for the energy contents in the universe.

It is useful to remind ourselves that, if we set $F(Q) = Q$, we have what is known as the Symmetric Teleparallel Equivalent of General Relativity (STEGR), an equivalent theory to GR which is based solely on non-metricity.

The non-metricity scalar is given by \cite{Jimenez2019}
\begin{equation}
    \label{eq:non-metricity-scalar}
    Q = -Q_{\alpha \mu \nu} P^{\alpha \mu \nu} \,,
\end{equation}
where $Q_{\alpha \mu \nu}$ is the non-metricity tensor given by

\begin{equation}
    Q_{\alpha \mu \nu} \equiv \nabla_{\alpha} g_{\mu \nu} \,.
\end{equation}
and $P^{\alpha}{}_{\mu \nu}$ is the non-metricity conjugate, which can be computed using the relation

\begin{equation}
    P^{\alpha}{}_{\mu \nu} = - \frac{1}{2} L^\alpha{}_{\mu \nu} + \frac{1}{4} (Q^\alpha - \tilde{Q}^\alpha) - \frac{1}{4} \delta^\alpha_{(\mu} Q_{\nu)} \,,
\end{equation}
where $L^\alpha{}_{\mu \nu}$ is known as the disformation tensor, which takes the form

\begin{equation}
    L^\alpha{}_{\mu \nu} = \frac{1}{2} Q^\alpha{}_{\mu \nu} - Q_{(\mu \nu)}{}^\alpha \,,
\end{equation}
being $Q_\alpha$ and $\tilde{Q}_\alpha$ two independent traces of the non-metricity, defined as

\begin{equation}
    Q_\alpha \equiv g^{\mu \nu} Q_{\alpha \mu \nu}, \quad \tilde{Q}_\alpha \equiv g^{\mu \nu} Q_{\mu \alpha \nu} \,.
\end{equation}

We restrict our analysis to a flat, homogeneous and isotropic universe, described by the FLRW metric

\begin{equation}
    ds^2 = -dt^2 + a^2(t)(dx^2 + dy^2 + dz^2) \,,
\end{equation}
where $t$ represents the cosmic time and $a(t)$ is the scale factor.

Under the previous assumption, the non-metricity scalar becomes

\begin{equation}
    Q = 6 H^2 \,,
\end{equation}
where $H = \dot{a}/a$ is the Hubble function.

Throughout this analysis we consider a universe permeated by a perfect fluid composed by matter, with density $\rho_m$, and a cosmological constant $\Lambda$.

The Friedmann equations under these assumptions are

\begin{gather}
    \label{eq:bg1}
    2 F_{Q} H^2 - \frac{1}{6}F = \frac{8 \pi G}{3}\rho_m + \frac{\Lambda}{3} \,, \\
    \label{eq:bg2}
    (12 F_{QQ} H^2 + F_{Q})\dot{H} = - 4 \pi G \rho_m \,.
\end{gather}
where the index $Q$ denotes a partial derivative with respect to $Q$.

As shown in \cite{Jimenez2019}, the most general function $F(Q)$ which replicates a $\Lambda$CDM cosmological background is given by

\begin{equation}
    \label{eq:model}
    F(Q) = Q + M \sqrt{Q} + C \,,
\end{equation}
where $M$ and $C$ are constants. 

The previous equation can be inserted back into equation \autoref{eq:bg1}, revealing that the Hubble function becomes

\begin{equation}
    H^2 = \frac{8 \pi G}{3} \rho_m + \frac{\Lambda}{3} + \frac{C}{6} \,.
\end{equation}

Notice that $C$ behaves the same as a cosmological constant. As such, without any loss of generality, we set $C = 0$. The Hubble function now reads

\begin{equation}
    H(z) = H_0 \sqrt{\Omega_m(1+z)^3 + 1-\Omega_m} \,,
\end{equation}
where $\Omega_m$ is the relative abundance for matter, $\Omega_\Lambda = 1 - \Omega_m$ the relative abundance for the cosmological constant and $H_0 = 100h \, \text{km} \, \text{s}^{-1} \text{Mpc}^{-1}$ is the Hubble constant.

Although the Hubble function is the same as if we were using $\Lambda$CDM, the additional free parameter induces a modification in the friction term in the equation of propagation of gravitational waves such that the gravitational waves luminosity distance becomes \cite{Jimenez2019,Belgacem2019}  

\begin{equation}
    \label{eq:dgw}
    d_\text{GW}(z) = \exp\left( \frac{1}{2} \int_0^z \frac{d}{d\ln a} ( \ln \, F_Q )\, \frac{d z'}{1+z'} \right) d_L(z) \,,
\end{equation}
where $d_L(z)$ is the electromagnetic luminosity distance given by 

\begin{equation}
    d_L(z) = (1+z) c \int_0^z \frac{1}{H(z')} dz' \,.
\end{equation}

Equation \autoref{eq:dgw} integrates to the usual result for modified gravity theories

\begin{equation}
   \label{eq:dlgw}
    \frac{d_\text{GW}(z)}{d_L(z)} = \sqrt{\frac{F_Q(0)}{F_Q(z)}} = \sqrt{\frac{G_{\text{eff}}(z)}{G_{\text{eff}}(0)}}
\end{equation}

where in the second equality we used 
\begin{equation}
    \label{eq:GenericGeff}
    G_{\text{eff}}/G = 1/F_Q \,.
\end{equation}

From equations \autoref{eq:model} and \autoref{eq:GenericGeff} we can obtain the specific form of the effective gravitational constant for this model, which reads

\begin{equation}
    \label{eq:Geff}
    G_{\text{eff}}(z) = \frac{G}{1 + M/2\sqrt{6}H(z)} \,.
\end{equation}

Using equations \autoref{eq:Geff} and \autoref{eq:dlgw}, the measured luminosity distance of a GW for this model is

\begin{equation}
    \label{eq:dlgw-final}
    d_{\text{GW}}(z) = \sqrt{\frac{2\sqrt{6} + M}{2\sqrt{6} + M/E(z)}} d_L(z) \,,
\end{equation}
where $M$ has been re-defined to be in units of $1/H_0$ and $E(z) \equiv H(z)/H_0$.

By looking at the GW luminosity distance present in equation \autoref{eq:dlgw-final}, one can see that there is a singularity at $M = -2\sqrt{6} E(z)$. To ensure that the value for the luminosity distance for GWs is strictly physical (i.e. a real positive number for all redshifts), we must require that $M$ must have a lower bound at $M = -2 \sqrt{6}$.

This cosmological model has also been addressed in \cite{Barros2020}, where it has been shown to alleviate the $\sigma_8$ tension and in \cite{Frusciante2021}, where further analysis using scalar angular power spectra, matter power and GWs propagation was carried out. Similar models were also briefly studied in \cite{Jimenez2017}. A more general model of the form $F(Q) = Q + \alpha Q^n$ was studied in \cite{Khyllep2021}, and a model of similar form was constrained using observational data in \cite{Ayuso2020, Atayde2021}. An $F(Q)$ model which does not feature dark energy is shown to be statistically similar to $\Lambda$CDM when confronted against observational data in \cite{Anagnostopoulos2021}.  A study of an $F(Q)$ model with an identical background evolution of the DGP models was developed in \cite{Ayuso2021}. Observational constraints on various different polynomial parametrizations of $F(Q)$ gravity as an explicit function of the redshift were studied in \cite{Lazkoz2019, Mandal2020}. An analysis of the linear perturbations for $F(Q)$ gravity using a designer approach was developed in \cite{Albuquerque2022}.

%% file: sections/data.tex
\section{Datasets}
\label{sec:datasets}

In order to constrain the three parameters for our model, $(h, \Omega_m, M)$, we will be using data from real type Ia Supernova (SNIa), as well as standard siren mock catalogs generated for a $\Lambda$CDM universe (\textit{i.e.} $M = 0$).

\subsection{Type Ia Supernova}
\label{subsec:snia}

When one looks at the correction introduced by our model in the luminosity distance of gravitational waves, as shown in equation \autoref{eq:dlgw-final}, one can see that there is a factor of $M/E(z)$, where $E(z)$ implicitly depends on $\Omega_m$. This causes a degeneracy between $M$ and $\Omega_m$, which prevents us from constraining both quantities using exclusively standard sirens. As such, we decided to fix the value of $\Omega_m$ using Type Ia Supernova (SNIa).

The dataset of SNIa considered is the Pantheon sample, which was developed in \cite{pantheon}, and is available in a public repository at \cite{pantheon-repo}. For performance reasons, the binned sample was used. 

Following \cite{Goliath2001} we performed a marginalization on $H_0$ and the absolute magnitude, such that the likelihood for the SNIa reads

\begin{equation}
    \label{eq:likelihoodsnia}
    L = \exp \left( -\frac{1}{2} \left( A - \frac{B^2}{C} \right) \right) \,,
\end{equation}
where A, B and C are given by

\begin{gather}
    A \equiv \sum_{i = 1}^n \frac{\Delta^2(z_i)}{\sigma^2(z_i)} \,, \\
    B \equiv \sum_{i = 1}^n \frac{\Delta(z_i)}{\sigma^2(z_i)} \,, \\
    C \equiv \sum_{i = 1}^n \frac{1}{\sigma^2(z_i)} \,,
\end{gather}
being $\sigma(z)$ the error for a SNIa measurement at redshift $z_i$ and

\begin{equation}
    \Delta (z_i) \equiv m^{\text{(obs)}}(z_i) - 5\log{\left( \frac{H_0}{c}d_L(z_i) \right)} \,,
\end{equation}
the difference between the observed magnitude $m^{\text{(obs)}}$ and the $H_0$ independent luminosity distance

\subsection{Standard Sirens}
\label{subsec:standard-sirens}

The single confirmed standard siren event detected so far, GW170817 \cite{GW170817}, and the possible detection of an EM counterpart to GW109521 \cite{GW190521, GW190521-EM}, are unable to distinguish between this model and $\Lambda$CDM. As such, we create standard sirens mock catalogs, which we use to forecast our model.

Here, we outline the procedure employed to generate such catalogs:

\begin{enumerate}
    \item Consider the theoretical distribution of standard sirens events, as a function of redshift, for a given observatory;
    \item Obtain a redshift $z_*$ by sampling the previous distribution; 
    \item Generate the corresponding luminosity distance $d_L(z_*)$ using $\Lambda$CDM, with values $\Omega_m = 0.284$, the most likely value given by the SNIa, and $h = 0.7$
    \item Consider the error $\sigma_\text{tot}$, as a function of redshift, for a given observatory;
    \item Compute the corresponding error for the obtained redshift $\sigma_\text{tot}(z_*)$, and consider it to be the $1 \sigma$ region for the luminosity distance;
    \item Sample from a Gaussian distribution with mean in $d_L(z_*)$ the standard deviation equal to $\sigma_\text{tot}(z_*)$, and consider the new value to be the observed value of the luminosity distance, for that redshift $d_L^{(\text{obs})}(z_*)$.
\end{enumerate}

The last step is used to ensure that we obtain a more realistic catalog, such that the mock value does not fall on top of the fiducial value.

For the previously generated standard sirens, we have decided to work with a Gaussian likelihood, which takes the form

\begin{equation}
    \label{eq:likelihood-gw}
    L = \prod_{i=1}^N \frac{1}{\sqrt{2 \pi} \sigma_\text{tot}(z_i)} \exp \left(- \frac{1}{2} \left[ \frac{d_\text{GW}^{\text{(obs)} }(z_i) - d_{\text{GW}}(z_i)}{\sigma_\text{tot}(z_i)} \right]^2 \right) \,,
\end{equation}
where $d^\text{(obs)}_\text{GW}(z)$ is the observed luminosity distance, which we generate with the procedure outlined before, $d_\text{GW}(z)$ is the theoretical gravitational wave luminosity distance given by equation \autoref{eq:dlgw-final}, and $N$ is the number of standard siren events.

We will now analyze each of the observatories we have considered throughout this analysis as sources of standard siren events.

\subsubsection{LIGO-Virgo Forecasts}
\label{subsubsec:ligo}

The theoretical distribution of standard siren events detected by LIGO-Virgo is obtained by taking the distance distribution found in \cite{lagos2019}. We sample it to obtain a value for the luminosity distance and assuming $\Lambda$CDM we compute the corresponding redshift. We then follow the steps 4, 5 and 6 of the procedure outlined before to obtain a mock catalog.

Following what was developed in \cite{Baker2021}, we consider that each catalog is composed of $N = 50$ events and that the error as a function of redshift is given by
\begin{equation}
    \label{eq:LIGO-error}
    \sigma^2_\text{LIGO-Virgo} = \sigma_{d_L}^2 + \left( \frac{d}{dz}(d_L) \sigma_\text{spect} \right)^2 \,,
\end{equation}
with

\begin{equation}
    \sigma_{d_L} = \frac{5.63 \times 10^{-4}}{\text{Mpc}} d_L^2(z) \,,
\end{equation}
being the component which provides a direct contribution to the luminosity distance error, where $d_L(z)$ is in units of Mpc, and

\begin{equation}
    \sigma_\text{spect} = 0.005(1+z) \,,
\end{equation}
is the error contribution for the redshift, due to spectroscopic measurements.

\subsubsection{LISA}
\label{subsubsec:LISA}

In \cite{Caprini2016} the authors presented the redshift distribution of standard siren events visible to LISA for three distinct populations of MBHBs: \textit{No Delay}, \textit{Delay} and \textit{Pop III}, and for three different mission specifications. We have chosen to work with the redshift distribution for the mission specification L6A2M5N2, which is the closest to the proposed mission specification \cite{LISA-proposal}.

\begin{figure}[h!]
    \centering
    \includegraphics[width=\columnwidth]{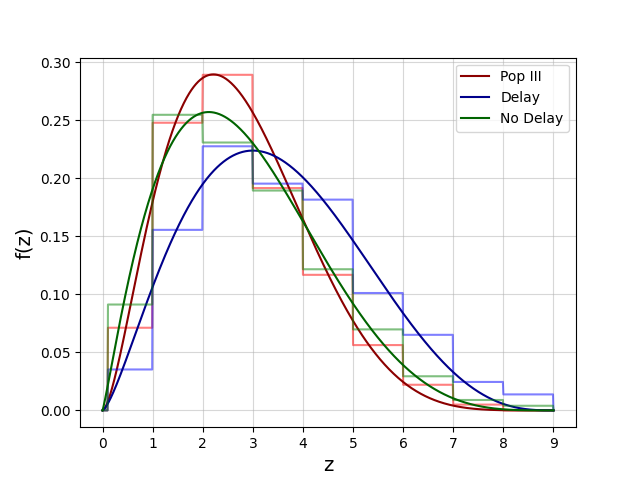}
    \caption{Normalized redshift distribution for LISA MBHB events, with visible EM counterpart, for the L6A2M5N2 mission specification, for populations Pop III, Delay and No Delay, fitted with equation \autoref{eq:beta} with the best fit values present in table \ref{tab:bestfit}.}
    \label{fig:mbhb-dist}
\end{figure}

As it was done in \cite{Speri2021}, we have modified the redshift distribution to include no events at $z < 0.1$. The normalized redshift probability distribution function for the three MBHB populations are presented in figure \autoref{fig:mbhb-dist}.

For the sake of convenience, we have fitted each of the histograms with a beta distribution, which takes the form

\begin{equation}
    \label{eq:beta}
    f(z) = \gamma \left( \frac{z}{9} \right)^{\alpha - 1} \left(1 - \frac{z}{9} \right)^{\beta - 1} \,,
\end{equation}
where the best fit values for each population are present in table \autoref{tab:bestfit}.

\begin{table}[h!]
\centering
\begin{tabular}{|c|c|c|c|}
\hline
         & $\alpha$ & $\beta$ & $\gamma$ \\ \hline
Pop III  & 2.64     & 6.03    & 11.95   \\ \hline
Delay    & 2.42     & 3.84    & 3.37    \\ \hline
No Delay & 2.14     & 4.7     & 3.61   \\ \hline
\end{tabular}
\caption{Best fit values for the redshift distribution for the Pop III, Delay and No Delay MBHB populations, considering equation \autoref{eq:beta}, and present in figure \ref{fig:mbhb-dist}.}
\label{tab:bestfit}
\end{table}

From \cite{Speri2021}, the estimated error as a function of redshift for LISA is given by

\begin{equation}
    \label{eq:LISA-error}
    \begin{aligned}
        \sigma^2_\text{LISA} = \sigma_{\text{delens}}^2 + \sigma_\text{v}^2 + \sigma_{\text{inst}}^2 + \left( \frac{d}{dz} (d_L) \sigma_{\text{photo}} \right)^2 \,,
    \end{aligned}
\end{equation}
where the total lensing contribution is given by

\begin{equation}
    \sigma_{\text{delens}} =  F_{\text{delens}} \, \sigma_{\text{lens}} \,,
\end{equation}
where

\begin{equation}
    \sigma_{\text{lens}} = 0.066 \left( \frac{1 - (1+z)^{-0.25}}{0.25} \right)^{1.8} d_L(z) \,,
\end{equation}
is the analytically estimated weak lensing contribution and

\begin{equation}
    F_{\text{delens}} = 1 - \frac{0.3}{\pi/2} \arctan{(z/0.073)} \,,
\end{equation}
the delensing factor, which includes the possibility of estimating the lensing magnification distribution and partially correct the weak lensing contribution.

Further contributions include the error coming from the peculiar velocity of the sources
\begin{equation}
    \sigma_\text{v} = \left[ 1 + \frac{c(1+z)^2}{H(z)d_L(z)} \right] \frac{500 \ \text{km/s}}{c} d_L(z) \,,
\end{equation}
the LISA instrumental error
    
\begin{equation}
    \sigma_{\text{inst}} = 0.05 \left( \frac{d_L^2(z)}{36.6 \text{Gpc}} \right) \,,
\end{equation}

and finally the redshift error associated with photometric measurements
\begin{equation}
    \sigma_{\text{photo}} = 0.03(1 + z), \text{ if $z > 2$} \,.
\end{equation}
We neglect the spectroscopic redshift error for redshifts below $z =2$.

As for the population, according to \cite{Speri2021}, the major difference is expected to come from the number of events detected. As such, we have decided to work with the \textit{No Delay} population, as it seems to provide a middle ground between the other two.

The most conservative estimate for the number of events using the current hardware specification, out of all the MBHB populations, according to \cite{Tamanini2017}, points towards $N = 15$ standard siren events detected by LISA, for the proposed mission lifetime of 4 years.

\subsubsection{ET}
\label{subsubsec:ET}

Following the procedure employed on \cite{Belgacem2018}, we expected that the ET will observe $N = 10^3$ BNS standard sirens events, over a three year period. The redshift probability distribution function distribution for the BNS is given by

\begin{equation}
    f(z) = \frac{4 \pi \mathcal{N} r(z) d_L^2(z)}{H(z)(1+z)^3} \,,
\end{equation}
where $\mathcal{N}$ is a normalization constant, set to be such that the following integral holds

\begin{equation}
    N = \int_{z_\text{min}}^{z_\text{max}} f(z) dz \,.
\end{equation}
where it is estimated that $z_\text{min} = 0.07$ is the minimum redshift at which we expect to observe standard siren events, and $z_\text{max} = 2$ the maximum redshift at which the ET can have meaningful observations.

The function $r(z)$ is the coalescence rate at redshift $z$, which is given by

\begin{equation}
  r(z) =
    \begin{cases}
      1+2z & \text{if } z \leq 1 \\
      (15-3z)/4 & \text{if } 1 < z < 5\\
      0 & \text{otherwise} \,.
    \end{cases}       
\end{equation}

The estimated error as a function of redshift for this observatory is given by

\begin{equation}
    \label{eq:ET-error}
    \sigma^2_\text{ET} = \sigma^2_\text{inst} + \sigma^2_\text{lens} \,,
\end{equation}
where

\begin{equation}
    \sigma_\text{inst} \approx (0.1449z - 0.0118z^2 + 0.0012z^3) \, d_L(z) \,,
\end{equation}
is the ET instrumental error and

\begin{equation}
    \sigma_\text{lens} \approx 0.05 z \, d_L(z) \,,
\end{equation}
is the estimated lensing contribution.
We neglect the error from the spectroscopic redshift measurements.

%% file: sections/methodology.tex
\section{Methodology}
\label{sec:methodology}

\subsection{Sampling}
\label{subsec:sampling}

The results shown here were obtained using PyStan \cite{pystan}, a Python interface to Stan \cite{stan}, a statistical programming language which implements the No-U-turn sampler, a variant of the Hamiltonian Monte Carlo. The output was then analyzed using GetDist \cite{GetDist}.

For each run we executed 5 independent chains, each with 2500 samples on the posterior distribution and 500 warm-up steps, where the initial values were sampled from a gaussian distribution with mean around the fiducial values, and a standard deviation of approximately 10\% of the corresponding mean for each parameter.

We restricted the values which the parameters could take such that the value of $M$ was set to be larger than $-2\sqrt{6}$, $\Omega_m$ to be between 0 and 1, and $h$ to have a lower bound in 0.

We then added weakly informative priors, such that $h \sim \text{normal}(0.7, 10)$, $\Omega_m \sim \text{normal}(0.284, 10)$ and $M \sim \text{normal}(0, 5)$.

The Stan model files, the output of the MCMC as well as the corresponding analysis is available in \cite{repo}.

\subsection{Catalog selection}
\label{subsec:catalog}

To provide statistical confidence that the catalogs used throughout this analysis represent a wide range of outcomes, we generated several catalogs and studied three possible cases: the best, the median and the worst. All of the generated datasets are available in \cite{repo}.

The criteria used to categorize each catalog was based on how small the 1$\sigma$ region for the parameter $M$ was. We could also have chosen to work with $\Delta h$, but this would yield similar results since we found that $\Delta h \propto \Delta M $.

For LISA we generated 15 different catalogs, of which the best, median and worst can be seen in figure \autoref{fig:LISA-9,10,12}.

\begin{figure}[h!]
    \centering
    \includegraphics[width=\columnwidth]{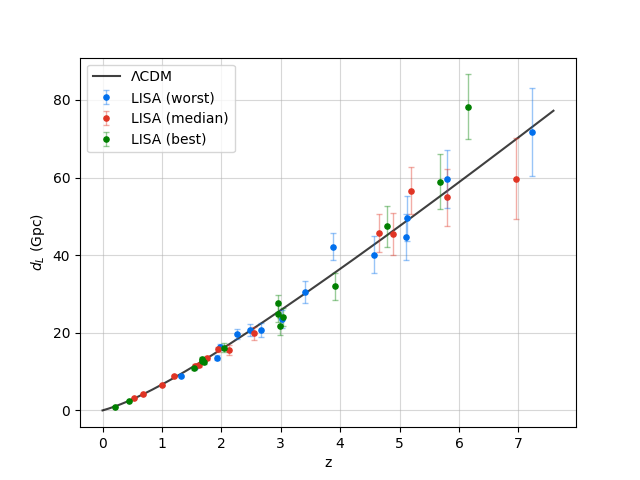}
    \caption{Luminosity distance, in Gpc, as a function of redshift for the worst, median and best LISA catalogs. The $\Lambda$CDM luminosity distance is plotted as a solid gray line.}
    \label{fig:LISA-9,10,12}
\end{figure}

\begin{figure}[h!]
    \centering
    \includegraphics[width=\columnwidth]{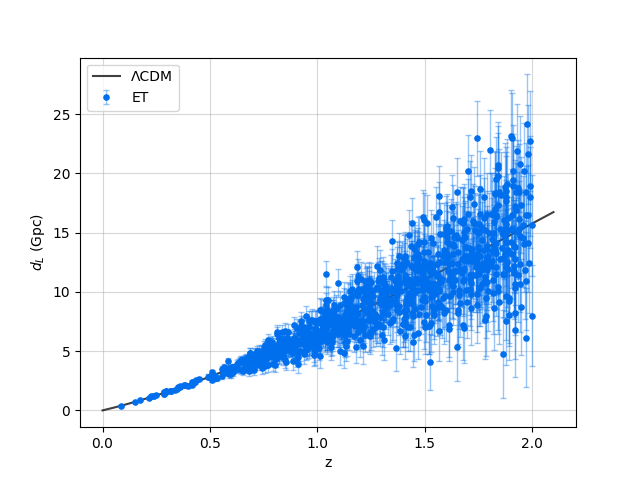}
    \caption{Luminosity distance, in Gpc, as a function of redshift for the ET catalog. The $\Lambda$CDM luminosity distance is plotted as a solid gray line.}
    \label{fig:ET-4}
\end{figure}

For the ET we generated 5 different catalogs, all of which showed similar results. This is something we already expected due to the large number of events per catalog. As such, in this paper, we will only show the results for the one presented in figure \autoref{fig:ET-4}. 

For LIGO-Virgo we generated 15 different catalogs, of which none of them could provide proper constrains. As such, we decided to categorize each LIGO-Virgo catalog based on how well it would complement the worst LISA catalog. The best, median and worst LIGO-Virgo catalogs, categorized using this modified criteria, are shown in figure \autoref{fig:LIGO-9,10,12}.

\begin{figure}[h!]
    \centering
    \includegraphics[width=\columnwidth]{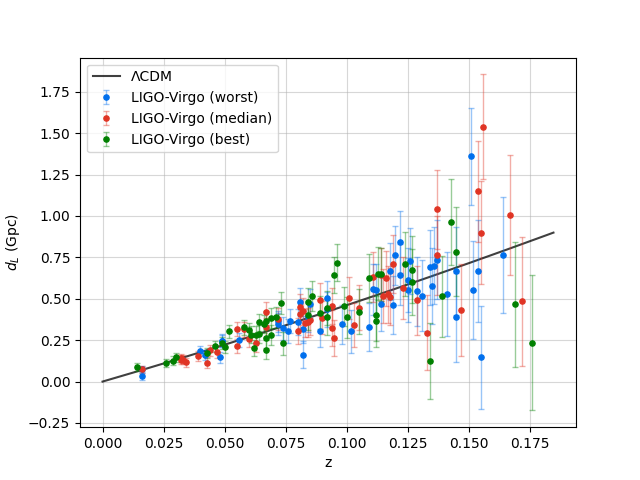}
    \caption{Luminosity distance, in Gpc, as a function of redshift for the worst, median and best LIGO-Virgo catalogs. The $\Lambda$CDM luminosity distance is plotted as a solid gray line.}
    \label{fig:LIGO-9,10,12}
\end{figure}

%% file: sections/results-discussion.tex
\section{Results and Discussion}
\label{sec:results-discussion}

Using the sampling method described in subsection \ref{subsec:sampling} and the catalogs present in subsection \ref{subsec:catalog}, we will now address the forecasts for our cosmological model.

Starting with the forecasts set by LISA, presented in figure \autoref{fig:FQ_LISA-9,10,12_SNIa-binned}, we can see that the best and median catalog show similar results, which contrasts with the worst catalog, that provides significantly worse results when compared to the other two.

\begin{figure}[h!]
    \centering
    \includegraphics[width=\columnwidth]{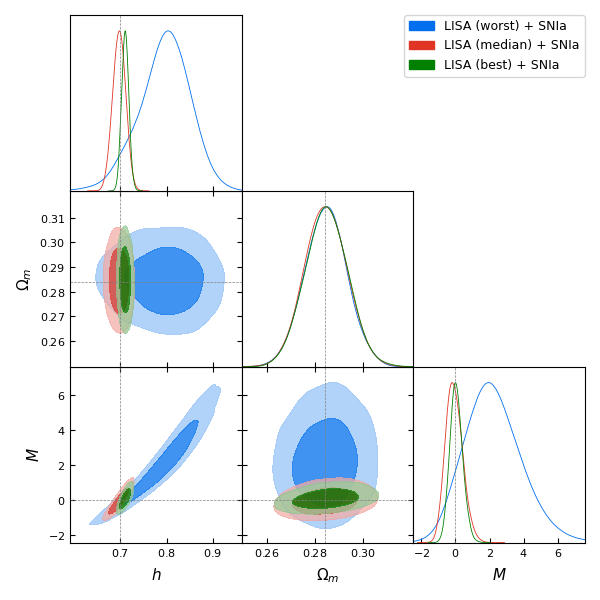}
    \caption{The 1$\sigma$ and 2$\sigma$ contours for the likelihood in the $(h, \Omega_m, M)$ plane for the model given by equation \autoref{eq:model}, with contribution from the worst, median and best LISA catalogs, with SNIa. Dotted lines represent the fiducial values $(h, \Omega_m, M) = (0.7, 0.284, 0)$.}
    \label{fig:FQ_LISA-9,10,12_SNIa-binned}
\end{figure}

To understand this difference, we refer back to figure \autoref{fig:LISA-9,10,12}, where the considered LISA catalogs are displayed. By comparing the catalogs with the corresponding forecasts we see a pattern: the best LISA catalog is the one which features more low redshift events, $0 \lesssim z \lesssim 6$, the median catalog has events with slightly larger redshifts, $0.5 \lesssim z \lesssim 7$, while the worst catalog favors higher redshift events, $1 \lesssim z \lesssim 7$. As such, we inferred that to better constrain this model, LISA will give best results when featuring plenty of low redshift events.

Considering that, out of the 15 generated LISA catalogs, only 2 provided similar error bars to the ones obtained in the worst LISA catalog, we expect the chances of obtaining a bad LISA catalog to be low.

\begin{figure}[h]
    \centering
    \includegraphics[width=\columnwidth]{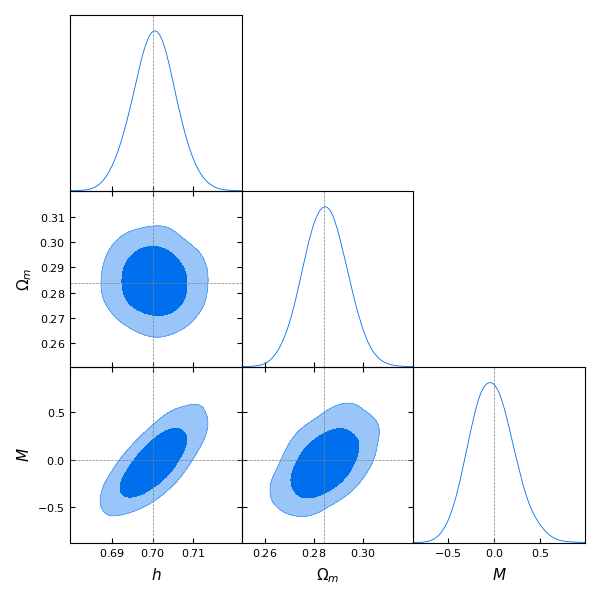}
    \caption{The 1$\sigma$ and 2$\sigma$ contours for the likelihood in the $(h, \Omega_m, M)$ plane for the model given by equation \autoref{eq:model}, with contribution from the ET, with SNIa. Dotted lines represent the fiducial values $(h, \Omega_m, M) = (0.7, 0.284, 0)$.}
    \label{fig:FQ_ET-4_SNIa-binned}
\end{figure}

The forecasts set by ET are presented in figure \autoref{fig:FQ_ET-4_SNIa-binned}. When we compare the ET results against those of LISA, we can see that the ET is able to provide a $1\sigma$ region approximately $65\%$ smaller than that of LISA alone. This goes in hand with our previous statement, that LISA should seek low redshift events. The range at which the ET is expected to operate, $z \in [0.07, 2]$, as shown in the catalog in figure \autoref{fig:ET-4}, is far lower than the redshift limits for LISA. Granted, the ET also provides a significant larger number of events when compared to LISA. 

These results show that, even though the luminosity distance for our model deviates from $\Lambda$CDM as the redshift increases, the error bars increase faster, in such a way that high redshift events are less useful to constrain this model. On the other hand, we note that very low redshift events are more sensitive to the peculiar velocities and consequently increase the error on the parameters. In addition, the luminosity distance in our model is practically indistinguishable from $\Lambda$CDM at low redshifts. It is possible to show that there is indeed a sweet spot around $z \sim 0.6$ for LISA and $z \sim 1$ for the ET.

LIGO-Virgo alone, as stated before, is not able to set any constrains on our parameters. To understand why, we refer to figure \autoref{fig:LISA-9,LIGO-13,ET-4,low-redshifts}, where we can see a comparison between the ET, LISA and LIGO-Virgo in the latter redshift observation band, $z \in [0, 0.2]$, showing that LIGO-Virgo has significantly larger error bars.

\begin{figure}[h!]
    \centering
    \includegraphics[width=\columnwidth]{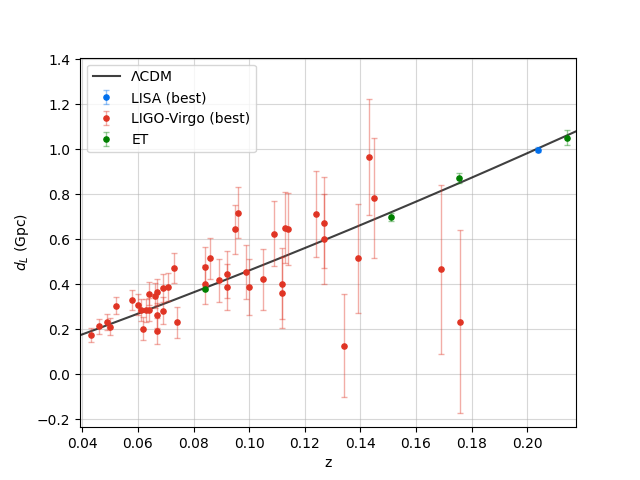}
    \caption{Luminosity distance, in Gpc, as a function of redshift for the best LISA catalog, the best LIGO-Virgo catalog and the ET catalog. The $\Lambda$CDM luminosity distance is plotted as a solid gray line.}
    \label{fig:LISA-9,LIGO-13,ET-4,low-redshifts}
\end{figure}

We categorized each LIGO-Virgo catalog based on how well it complements the worst LISA catalog because the best and the median LISA catalogs do not show significant improvements when used in conjunction with any of the LIGO-Virgo catalogs. The resulting constrains for LIGO-Virgo are presented in figure \autoref{fig:FQ_LISA-12_SNIa-binned_none,LIGO-1,2,13}.

\begin{figure}[h!]
    \centering
    \includegraphics[width=\columnwidth]{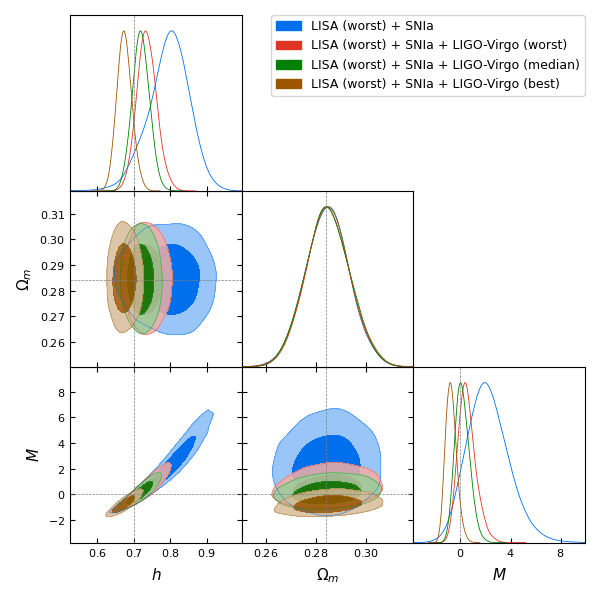}
    \caption{The 1$\sigma$ and 2$\sigma$ contours for the likelihood in the $(h, \Omega_m, M)$ plane for the model given by equation \autoref{eq:model}, with contribution from the worst LISA catalog, SNIa and the worst, median and best LIGO-Virgo catalog. Dotted lines represent the fiducial values $(h, \Omega_m, M) = (0.7, 0.284, 0)$.}
    \label{fig:FQ_LISA-12_SNIa-binned_none,LIGO-1,2,13}
\end{figure}

These results reveal that even if future data resembles a worst LISA catalog, LIGO-Virgo is expected to significantly increase the quality of the constraints.

When we add each of the LISA and LIGO-Virgo catalogs to the single ET catalog, where the corner plot is not shown here for brevity, only in the case of the best LISA catalog there are visible improvements. This is expected, given the large number of events measured by the ET when compared to LISA and LIGO-Virgo, and that LIGO-Virgo features much larger error bars for the same redshift region.

A quantitative summary of the more relevant results we have obtained in our analysis can be seen in table \autoref{tab:summarize-constrains}, where we show the 1$\sigma$ region for $M$, as well as the relative size of the 1$\sigma$ region for $M$ when compared to the base case, the best LISA catalog together with the ET.

\begin{table}[h!]
\centering
\begin{tabular}{|c|c|c|}
\hline
\textbf{Catalog}                             & $\mathbf{\Delta M}$ & \textbf{Relative Size} \\ \hline
LISA (best) + ET                             & 0.19       & 1             \\ \hline
ET                                           & 0.24       & 1.26          \\ \hline
LISA (best)                                  & 0.37       & 1.94          \\ \hline
LISA (worst) + LIGO-Virgo (best)             & 0.44       & 2.31          \\ \hline
LISA (median)                                & 0.50       & 2.63          \\ \hline
LISA (worst)                                 & 1.65       & 8.68          \\ \hline
\end{tabular}
\caption{Summary of the 1$\sigma$ region for $M$, for the relevant datasets, as well as its relative size using the best LISA catalog with the ET as the base case.}
\label{tab:summarize-constrains}
\end{table}

%% file: sections/conclusions.tex
\section{Conclusions}
\label{sec:conclusions}

We considered a model of Symmetric Teleparallel Gravity that mimics a $\Lambda$CDM background, with one additional free parameter, $M$. Departures from $\Lambda$CDM only arise at the perturbative level.

We have described the procedure used to generate standard sirens mock catalogs for LISA, the ET and LIGO-Virgo.

For LISA, we studied three representative catalogs: the best, median and the worst, each one consisting of 15 standard sirens events. We obtained the best LISA catalog when most of the events are present at low redshifts, while the worst catalog is composed mostly of high redshift events. According to our analysis, the worst LISA catalog is unlikely to be observed.

For the ET we observed that all catalogs provide similar constrains and as such only one was considered. This is due to each featuring 1000 standard sirens events, a number large enough to represent the underlying probability distribution function.

Finally for LIGO-Virgo, our analysis showed that even with a set of 50 standard sirens, these are not enough to provide meaningful constrains. Instead, we categorized each LIGO-Virgo catalog based on how well it would complement the worst LISA catalog. We then studied three representative scenarios: the best, median and worst LIGO-Virgo catalogs. We showed that one can rely on future events obtained by LIGO-Virgo to improve the quality of the constrains set by the worst LISA catalog. By contrast, for the best or the median LISA catalogs, none of the LIGO-Virgo catalogs made significant improvements.

We can improve the constrains set by the ET using the best LISA catalog, whereas the median and the worst LISA catalog made no significant improvements. Likewise, none of the LIGO-Virgo catalogs made any improvements whatsoever to the constrains set by the ET alone.

Even though our model diverges from $\Lambda$CDM as the redshift increases, the error bars for all observatories increase faster, making high redshift observations less useful to provide constrains.

We quantified the 1$\sigma$ region for $M$, the various scenarios which are consistent with previous estimates using a simple parametrization in $F(Q)$. We note that we fixed $\Omega_{\rm m}$ using supernovae and we could further constrain the cosmology by adding other observables such as CMB. 

%% file: sections/acknowledgments.tex
\begin{acknowledgments}
The authors acknowledge support from Funda\c{c}\~ao para a Ci\^encia e a Tecnologia via the following projects: 
UIDB/04434/2020 \& UIDP/04434/2020,
CERN/FIS-PAR/0037/2019,
PTDC/FIS-OUT/29048/2017,
COMPETE2020: POCI-01-0145-FEDER-028987 \& FCT: PTDC/FIS-AST/28987/2017, PTDC/FIS-AST/0054/2021, EXPL/FIS-AST/1368/2021.
\end{acknowledgments}